\documentclass[journal]{IEEEtran}
\usepackage{amsmath}
\usepackage{cite}
\usepackage{graphicx}
\usepackage{epstopdf}
\usepackage{amsfonts,amsmath,amssymb}
\usepackage{graphicx}
\usepackage{url}
\usepackage{bm}
\usepackage{bbm}
\usepackage{subfigure}
\usepackage{stfloats}
\usepackage{citesort}
\usepackage{color}
\usepackage{balance}
\usepackage{multirow}
\usepackage{balance}
\usepackage{makecell}
\newcommand{\tabincell}[2]{\begin{tabular}{@{}#1@{}}#2\end{tabular}}

\usepackage{algorithmic}
\usepackage{algorithm}

\DeclareMathOperator*{\argmax}{argmax}

\newtheorem{theorem}{\textbf{Theorem}}

\newtheorem{proposition}{\textbf{Proposition}}

\newtheorem{lemma}{\textbf{Lemma}}

\newcommand{\Hnull}{\mathcal{H}_0}
\newcommand{\Halt}{\mathcal{H}_1}

\newcommand{\Honull}{\mathcal{{D}}_0}
\newcommand{\Hoalt}{\mathcal{{D}}_1}

\begin{document}

\title{Optimal Transmit Power and Flying Location for UAV Covert Wireless Communications}
\author{Shihao Yan,  \IEEEmembership{Member, IEEE,} Stephen V. Hanly, \IEEEmembership{Fellow, IEEE,} and Iain B. Collings, \IEEEmembership{Fellow, IEEE}
\thanks{S. Yan, S. V. Hanly, and I. B. Collings are with the School of Engineering, Macquarie University, Sydney, NSW, Australia (Emails: \{shihao.yan, stephen.hanly, iain.collings\}@mq.edu.au).}
\thanks{Part of this work has been presented in IEEE ICC 2019\cite{yan2019hiding}.}
}


\maketitle

\vspace{-2.2cm}

\begin{abstract}
This paper jointly optimizes the flying location and wireless communication transmit power for an unmanned aerial vehicle (UAV) conducting covert operations. This is motivated by application scenarios such as military ground surveillance from airborne platforms, where it is vital for a UAV's signal transmission to be undetectable by those within the surveillance region. Specifically, we maximize the communication quality to a legitimate ground receiver outside the surveillance region, subject to: a covertness constraint, a maximum transmit power constraint, and a physical location constraint determined by the required surveillance quality. We provide an explicit solution to the optimization problem for one of the most practical constraint combinations. For other constraint combinations, we determine feasible regions for flight, that can then be searched to establish the UAV's optimal location. In many cases, the 2-dimensional optimal location is achieved by a 1-dimensional search. We discuss two heuristic approaches to UAV placement, and show that in some cases they are able to achieve close to optimal, but that in other cases significant gains can be achieved by employing our developed solutions.
\end{abstract}


\section{Introduction}

As a wide range of applications of Unmanned Aerial Vehicle (UAV) networks are emerging (e.g., military surveillance, forest fire detection, vehicle traffic control), ever more demanding requirements are being placed on their communication systems~\cite{valavanis2015handbook,zeng2016wireless}. Recent advances have included an optimization framework of user scheduling and association, power control, and trajectory design \cite{JointWu2018}, as well as altitude optimisation for UAV communication platforms aiming to maximise radio coverage on the ground \cite{hourani2014optimal}. Practical issues, including user location uncertainty, wind speed uncertainty, and polygonal no-fly zones have been considered in optimizing the resource allocation and trajectory for multiuser UAV communications~\cite{xu2020multiuser}.
The fundamental rate limits have been explored for UAV-enabled multiple access channels, where multiple ground users transmit individual information to a mobile UAV and where the UAV's trajectory is optimized to maximize the average sum-rate of all the users \cite{li2020fundamental}.

In this paper we consider UAV applications that require covert communications. We are motivated by application scenarios such as military ground surveillance from airborne platforms, where it is vital for a UAV's signal transmission to be undetectable by those within the surveillance region. Specifically, we focus on the problem of jointly optimizing the UAV's location and transmit signal power to maximize the communication quality to a legitimate ground receiver outside the surveillance region, subject to: a covertness constraint, a maximum transmit power constraint, and a physical location constraint determined by the required surveillance quality.


Secure communications have been considered for UAV networks in non-covert applications, including the possibility of enhancing physical layer security (e.g., \cite{ImprovWang2017,Securezhang2017,SecrecyZhou2017,An2018UAV,zhou2018improving,zhou2019UAV,cai2020joint}). In \cite{ImprovWang2017} a UAV-enabled mobile relay was proposed to improve physical layer security, where the UAV served as a mobile relay and optimized its location in order to enhance wireless communication security. In \cite{Securezhang2017} it was shown that a UAV can increase the achieved secrecy rate with an optimal flight trajectory, by simultaneously improving the channel quality to a legitimate receiver while reducing the channel quality to an eavesdropper. In \cite{An2018UAV} a UAV was used as an external jammer to transmit artificial noise, aiming to prevent an eavesdropper from eavesdropping on confidential information, where the joint optimization of the UAV's trajectory and transmit power was considered. A similar application scenario was also investigated in \cite{SecrecyZhou2017,zhou2018improving,cai2020joint}, where the secrecy outage probability, jamming coverage, and joint trajectory were considered in order to show the effectiveness of using a UAV-jammer to enhance physical layer security. None of these papers addressed covertness of communications.

To operate a UAV in a surveillance scenario, it is necessary to shield the very existence of its wireless transmissions, used for conveying surveillance results back to ground stations. Covert communication technology focuses on 
hiding the wireless transmission from a legitimate transmitter, commonly called Alice, to an intended receiver, Bob, in the presence of a warden Willie, who is trying to detect this covert transmission \cite{bash2015hiding,liu2018covert,yan2019lowpro}. The fundamental limit of covert communication was established in \cite{bash2013limits}, which was extended into different scenarios by considering various constraints and practical issues, including noise uncertainty \cite{he2017on}, uninformed jammer \cite{sobers2017covert}, relay networks \cite{hu2018covert,wang2019covertcom}, generative adversarial networks \cite{liao2020generative}, full-duplex technique \cite{Khurram2018fullduplex,shu2019delay}, and multiple antennas \cite{zheng2019multi,lu2019proactive}.

Most recently, covert communication has been considered for UAV networks. In \cite{Zhou2018Joint}, a UAV's trajectory and transmit power were jointly optimized in order to aid it transmitting critical information to a legitimate ground user without being detected by Willie. In \cite{wang2020secrecyandcovert} the UAV was acting as the warden Willie and a multi-hop relaying strategy (e.g., the number of hops, transmit power) was optimized to maximize the throughput of a ground network subject to a covertness constraint. The horizontal trajectory (with a fixed height) of a UAV was tackled in \cite{Zhou2018Joint}, where multiple ideal assumptions, e.g., infinite symbols in each flying time slot and  line-of-sight (LoS) channels, were adopted.

In this work, we address the covertness of UAV wireless transmission in a scenario where the UAV conducts surveillance over a specific area and has to urgently transmit critical information back to a legitimate user covertly. We assume that Willie is in the surveillance area and the UAV flies in the vertical plane determined by the locations of Bob and Willie. We tackle the joint optimization of the UAV's location (distance and angle relative to Willie) and transmit power in order to maximize the communication quality at Bob subject to: 1) a covertness constraint (i.e., the total error rate at Willie is no less than a specific value), 2) a maximum transmit power constraint at the UAV, 3) a lower bound on the UAV's angle to Willie, and 4) a lower bound and an upper bound on the UAV's distance to Willie. The lower bound on the distance is to prevent the UAV getting too close to Willie where it could be observed. The lower bound on the angle is to ensure that the UAV gets a good enough view of the surveillance area near Willie, and the upper bound on the distance is to ensure sufficient image resolution.

Our main contributions are as follows:
\begin{itemize}
  \item We identify six scenarios that can arise due to the possible parameter and constraint combinations, and in each scenario provide expressions that significantly reduce the feasible region search space in order to find the UAV's optimal location.
  \item We explicitly show that in some scenarios, the UAV's 2-dimensional optimal location can be achieved by a 1-dimensional search.
  \item In the scenario where the distance between Bob and Willie is large relative to the lower and upper bounds on the UAV's distance to Willie, the most common scenario in practice, we explicitly determine the UAV's optimal location and transmit power. We show that the UAV's nearest feasible location to Bob (which naturally provides the strongest signal to Bob when not blocked) is not always the optimal location for covert communications.
  \item In the special case where the UAV is constrained to only fly directly above the surveillance area we solve the optimization problem explicitly. We show that the achieved effective signal-to-noise ratio (SNR) at Bob increases with the upper limit on the UAV’s height.
\end{itemize}

The reminder of this paper is organized as follows: Section~II details our considered system model with the adopted assumptions, Section~III considers the 2-dimensional location for the UAV, where six different scenarios are examined, in Section~IV we consider the vertical UAV case, and numerical results are presented in Section~V. In Section VI we draw conclusions.


\begin{figure}[!t]
    \begin{center}
        \includegraphics[width=0.8\columnwidth]{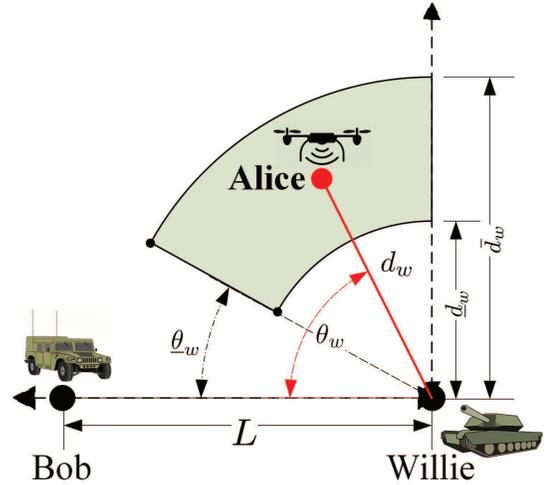}
        \caption{Illustration of the system model for delay-constrained covert communications with UAV.}
        \label{fig:fig1}
    \end{center}
\end{figure}

\section{System Model}\label{system_model}

\subsection{Channel Model}

The scenario of interest for covert communications in the context of UAV networks is illustrated in Fig.~\ref{fig:fig1}, where each of the UAV (i.e., Alice), Bob, and Willie is equipped with a single antenna. In this scenario, the UAV tries to transmit critical information to Bob covertly, while Willie is trying to detect this covert transmission to determine the presence of the UAV. Specifically, Alice transmits $n$ symbols $x[i]$ ($i = 1, 2, \dots, n$) to Bob, while Willie is collecting the corresponding observations on Alice's transmission to detect whether or not a wireless transmission took place. In this work, we consider a practical scenario, where  the transmission from Alice to Bob is subject to a delay constraint (the UAV collects some urgent information on the surveillance object and has to transmit this information in a timely manner to Bob), i.e., the transmitted signals are constrained by a maximum blocklength $n$. We denote the additive while Gaussian noise (AWGN) at Bob and Willie as $\omega_b[i]$ and $\omega_w[i]$, respectively, where $\omega_b[i] \sim \mathcal{N}(0,\sigma_b^2)$ and $\omega_w[i] \sim \mathcal{N}(0,\sigma_w^2)$, while $\sigma_b^2$ and $\sigma_w^2$ are the noise variances at Bob and Willie, respectively. In addition, we assume that $x[i]$, $\omega_b[i]$, and $\omega_w[i]$ are mutually independent. The transmit power of the UAV is denoted as $P$, i.e., we have $\mathbb{E}\{|x[i]|^2\} = P$. We also consider a maximum transmit power constraint $P \leq P_m$, where $P_m$ is the UAV's maximum transmit power.

The channels from the UAV to Bob and Willie are modelled as air-to-ground wireless channels with non-line-of-sight (NLoS) and probabilistic LoS components. Specifically, following \cite{hourani2014optimal} the path loss between the UAV and ground users (i.e., Bob and Willie) for the LoS and NLoS components are given by
\begin{eqnarray}\label{ag_model}
\rho_{j} = \left\{ \begin{aligned}\label{ncon}
        \ & d_j^{\xi_{\textrm{L}}}, ~~\text{for LoS component}\\
        \ & d_j^{\xi_{\textrm{N}}}, ~~\text{for NLoS component},
         \end{aligned} \right.
\end{eqnarray}
where $j \in \{b, w\}$, i.e., $d_j$ denotes the distance from the UAV to Bob or Willie, $\xi_{\textrm{L}}$ is the path loss exponent for the LoS channel component, and $\xi_{\textrm{N}}$ is the path loss exponent for the NLoS channel component. As per \cite{hourani2014optimal}, the LoS component does not always exist in the air-to-ground channels and the probability to have the LoS component in these channels depends on the environment parameters and the angle from the UAV to the ground user. The considered scenario is shown in Fig.~\ref{fig:fig1}, where $L$ is the distance between Bob and Willie.
Considering the above specific scenario, the probability to have the LoS component in the channels from the UAV to Bob or Willie is given by
\begin{align}\label{probability_LOS}
p_j = \frac{1}{1 + a \exp(-b[\theta_j - a])},
\end{align}
where $a$ and $b$ are the S-curve parameters that depend on the communication environment and $\theta_j$ ($j \in \{b, w\}$) is the elevation angle (in degrees) of the UAV (Alice) relative to Bob or Willie. Following many existing works (e.g., \cite{xu2020multiuser,Zhou2018Joint,wang2020secrecyandcovert,xie2020common}) in the literature on UAV networks, we approximate the air-to-ground wireless channels as LoS channels with Willie's path loss given by $d_w^{\xi_{{\textrm{L}}}}p_w +  d_w^{\xi_{{\textrm{N}}}}$, where $d_w^{\xi_{{\textrm{L}}}}p_w$ is the effective path loss for LoS components and $d_w^{\xi_{{\textrm{N}}}}$ is the path loss for the NLoS components. We note that we do not consider these channels as probabilistic channels, since the resultant Gaussian mixture likelihood functions lead to analytically intractable expressions for the covertness constraints.

\subsection{Binary Hypothesis Testing at Willie}

In order to detect the UAV's covert transmission, Willie has to distinguish between the following two hypotheses:
\begin{eqnarray}\label{hypothses}
 \left\{ \begin{aligned}\label{ncon}
        \ & \Hnull:~y_w[i] = \omega_w[i], \;\;\\
        \ & \Halt:~y_w[i] = g x[i] + \omega_w[i],
         \end{aligned} \right.
\end{eqnarray}
where $\Hnull$ denotes the null hypothesis where the UAV did not transmit signals, $\Halt$ denotes the alternative hypothesis where the UAV transmitted, $y_w[i]$ is the received signal at Willie for the $i$-th channel use, and $g$ denotes the air-to-ground channel from the UAV to Willie.

In this work, we use the total error rate to measure Willie's detection performance, which is given by
\begin{align}
\xi = \alpha + \beta,
\end{align}
where $\alpha \triangleq \Pr(\Hoalt|\Hnull)$ is the false positive rate, $\beta \triangleq \Pr(\Honull|\Halt)$ is the miss detection rate, while $\Hoalt$ and $\Honull$ denote the binary decisions as to whether the UAV's transmission occurred or not, respectively. We note that Willie's goal is to detect the presence of the UAV's transmission with the minimum total error rate $\xi^{\ast}$, which is achieved by Willie's optimal detector that minimizes $\xi$. Then, the covertness constraint can be written as $\xi^{\ast} \geq 1 - \epsilon$, where $\epsilon$ is a small value to determine the specific required covertness\cite{bash2013limits}.

The optimal test that minimizes $\xi$ at Willie is the likelihood ratio test given by
\begin{equation}\label{LRT}
\frac{\mathbb{P}_1 \triangleq \prod_{i = 1}^n f\left(y_w[i]|\Halt\right)}{\mathbb{P}_0 \triangleq \prod_{i = 1}^n f\left(y_w[i]|\Hnull\right)} \begin{array}{c}
\overset{\Hoalt}{\geq} \\
\underset{\Honull}{<}
\end{array}%
1,
\end{equation}
where $f(y_w[i]|\Hnull)$ and $f(y_w[i]|\Halt)$ are the likelihood functions of $y_w[i]$ under $\Hnull$ and $\Halt$, respectively, while $\mathbb{P}_0$ and $\mathbb{P}_1$ are the likelihood function of $\mathbf{y}_w$ under $\Hnull$ and $\Halt$, respectively. Following \eqref{hypothses}, we have
\begin{align}
f(y_w[i]|\Hnull) &= \mathcal{N}(0, \sigma_w^2).\label{likelihood0}
\end{align}
In this work, considering that the path loss of the NLoS component is much larger than that of the LoS component in a channel, under $\Halt$ we approximate the channel from Alice to Willie effectively as LOS and thus following \eqref{hypothses} we have
\begin{align}
f(y_w[i]|\Halt) &= \mathcal{N}(0, \bar{P}+\sigma_w^2),\label{likelihood1}
\end{align}
where $\bar{P} = P d_w^{\xi_{{\textrm{L}}}}p_w + P d_w^{\xi_{{\textrm{N}}}}$ is the effective received power at Willie if the UAV's wireless transmission is taking place. The optimal test at Willie can be constructed as per \eqref{LRT}, based on which the detection performance at Willie in terms of the minimum total error rate can be derived~\cite{yan2018delay}. However, we note that incomplete gamma functions are involved in the expression for the minimum total error rate \cite{yan2018delay}, which is difficult in further analysis. As such, in the following we present a lower bound on this minimum total error rate in terms of the KL divergence, which is given by \cite{bash2013limits,yan2018gaussian}
\begin{align}\label{lower_bound10}
\xi^{\ast} \geq 1 - \sqrt{\frac{1}{2}\mathcal{D}_{01}},
\end{align}
where $\mathcal{D}_{01}$ is the KL divergence from  $\mathbb{P}_0$ to $\mathbb{P}_1$ given by
\begin{align}\label{D01_ex}
\mathcal{D}_{01} = \frac{n}{2} \left[\ln\left(\frac{\bar{P}+\sigma_w^2}{\sigma_w^2}\right)-\frac{\bar{P}}{\bar{P}+\sigma_w^2}\right].
\end{align}
We note that the KL divergence from  $\mathbb{P}_1$ to $\mathbb{P}_0$ can also be used to achieve another lower bound on $\xi$. However, as shown in \cite{yan2018gaussian} the lower bound based on  $\mathcal{D}_{01}$ is tighter in our considered system model. As such, following \eqref{lower_bound10} we employ the following covertness constraint in this work
\begin{align}
\mathcal{D}_{01} \leq 2 \epsilon^2,
\end{align}
as it is sufficient to ensure $\xi^{\ast} \geq 1 - \epsilon$. Therefore, in the remaining part of this work we use $\mathcal{D}_{01} \leq 2 \epsilon^2$ as the covertness constraint.

\subsection{Delay-Constrained Transmission from the UAV to Bob}

The received signal at Bob for each channel use is
\begin{align}\label{xn}
y_b[i] = h x[i] + \omega_b[i],
\end{align}
where $h$ is the air-to-ground channel from the UAV to Bob. Following \cite{hourani2014optimal} and considering that the path loss for NLoS component in the air-to-ground channel is much larger than that of the LoS component, in this work we approximate the SNR at Bob effectively as
\begin{align}\label{effective_SNR}
\gamma_b(P, d_w, \theta_w) = \frac{P d_b^{\xi_{{\textrm{L}}}} p_b}{\sigma_b^2}=  \frac{Pf(d_w, \theta_w)}{\sigma_b^2},
\end{align}
where $f(d_w, \theta_w)$ is given by
\begin{align}\label{f_dw_thetaw}
&f(d_w, \theta_w) =\notag\\
 &\frac{(L^2 + d_w^2 - 2 d_w L \cos (\theta_w))^{\xi_{{\textrm{L}}}/2}}{1 + a \exp(-b[\frac{180}{\pi}\arcsin\left(\frac{d_w \sin (\theta_w)}{\sqrt{L^2 + d_w^2 - 2 d_w L \cos (\theta_w)}}\right) - a])}.
\end{align}
This is due to the fact that, following the geometry given in Fig.~\ref{fig:fig1}, we have $d_b^2 = L^2 + d_w^2 - 2 d_w L \cos (\theta_w)$ and $\theta_b = \arcsin\left({d_w \sin (\theta_w)}/{\sqrt{L^2 + d_w^2 - 2 d_w L \cos (\theta_w)}}\right)$. We recall that $P$ is the UAV's transmit power, $d_b^{\xi_{{\textrm{L}}}}$ is the path loss from the UAV to Bob, and $p_b$ is the probability to have a LoS component in the channel from the UAV to Bob. Here, we do not consider the NLoS components in Bob's channel from the perspective of a worst-case scenario, where a lower bound on Bob's SNR can be obtained.

For a fixed transmission rate at the UAV, the decoding error probability at Bob is not negligible when $n$ is finite and small~\cite{yan2018delay}. As such, the effective throughput, which quantifies the amount of information that can be transmitted reliably from the UAV to Bob, is a reasonable performance metric for the delay-constrained communication scenarios. As proved in \cite{sun2018short-packet}, this effective throughput is a monotonically increasing function of the corresponding SNR. As such, in this work we use the effective SNR given in \eqref{effective_SNR} as the performance metric for the delay-constrained transmission from the UAV to Bob.

\section{Optimal Location and Transmit Power for A 2-Dimensional UAV}\label{section_Vertical}

 In the considered scenario, the UAV has to determine its optimal location and transmit power in order to maximize the covert communication performance. Specifically, the optimization problem at the UAV is given by
\begin{subequations}\label{opt_2D}
\begin{align}
 \argmax_{P, d_w, \theta_w} ~~&\gamma_b(P, d_w, \theta_w),\\
~~~~\text{s.t.} ~~& \mathcal{D}_{01} \leq 2 \epsilon^2, \label{covertness_con2}\\
&P \leq P_m, \label{power_max_con}\\
&\underline{d}_w \leq d_w \leq \overline{d}_w, \label{distance_con2} \\
&\underline{\theta}_w \leq \theta_w \leq \frac{\pi}{2},\label{angle_con}
\end{align}
\end{subequations}
where \eqref{distance_con2} and \eqref{angle_con} jointly determine the constraints on the surveillance quality and the UAV operation scenario. Specifically, $d_w \leq \overline{d}_w$ guarantees that the UAV cannot be too far from Willie and $\underline{\theta}_w \leq \theta_w$ ensures that the UAV has an acceptable angle to Willie, which jointly guarantee a good surveillance quality. The constraint $\underline{d}_w \leq d_w$ 
is to avoid the UAV being visually observed by Willie. The constraint $\theta_w \leq \frac{\pi}{2}$ is due to the fact that the UAV will stay on the left side (where Bob locates) of Willie as shown in Fig.~\ref{fig:fig1}, rather than staying on the other side of Willie.

\subsection{Optimal Location without Constraints}

In order to solve the optimization problem \eqref{opt_2D}, we first tackle the optimal $d_w$ that maximizes $\gamma_b(P, d_w, \theta_w)$ for a given $\theta_w$ and the optimal $\theta_w$ that maximizes $\gamma_b(P, d_w, \theta_w)$ for a given $d_w$. We note that, if $d_w$ and $\theta_w$ are jointly optimized, the UAV's optimal location that maximizes $\gamma_b(P, d_w, \theta_w)$ would be at Bob's location. In this subsection, we identify some properties on the optimal $d_w$ for a given $\theta_w$ and the optimal $\theta_w$ for a given $d_w$, which will be used to facilitate solving the original optimization problem \eqref{opt_2D}.

We first focus on the optimal $d_w$ that maximizes $\gamma_b(P, d_w, \theta_w)$ for a given $\theta_w$ without any constraint. We recall that we have $\gamma_b(P, d_w, \theta_w) = Pf(d_w, \theta_w)/\sigma_b^2$, and thus maximizing $\gamma_b(P, d_w, \theta_w) $ is equivalent to maximizing $f(d_w, \theta_w)$.  As such, the corresponding optimization problem is given by
\begin{align}\label{opt_dw_nocovert}
 \argmax_{d_w} ~~&f(d_w, \theta_w),
\end{align}
where $\theta_w$ is fixed within $[0, \frac{\pi}{2}]$. We have the following lemma detailing the solution to the optimization problem \eqref{opt_dw_nocovert}.

\begin{lemma}\label{lemma1}
The solution to the optimization problem \eqref{opt_dw_nocovert}, i.e., the optimal value of $d_w$ (denoted as $d_w^\circ(\theta_w)$) that maximizes $f(d_w, \theta_w)$ for a given $\theta_w$, is unique and is given by the unique solution in $d_w$ to the equation \eqref{optimal_dw_solution}
\begin{align}\label{optimal_dw_solution}
\xi_{{\textrm{L}}} \left[d_w \!\!-\!\! L\cos(\theta_w)\right] \!+\!\frac{180 b L \sin(\theta_w)}{\pi}\left[1\!\!-\!\! p_b(d_w, \theta_w)\right] \!=\! 0,
\end{align}
where $d_w^\circ(\theta_w)$ lies in the closed interval $[L\cos(\theta_w), L/\cos(\theta_w)]$, ${\partial f(d_w, \theta_w)}/{\partial d_w} > 0$ for $d_w \leq L\cos(\theta_w)$, and
${\partial f(d_w, \theta_w)}/{\partial d_w} < 0$ for $ L/\cos(\theta_w) \leq d_w.$
\end{lemma}
\begin{IEEEproof}
The detailed proof is presented in Appendix A.
\end{IEEEproof}

 Intuitively, the results in Lemma~\ref{lemma1} are due to the fact that, for a fixed $\theta_w < 90^\circ$, as the UAV moves away from Willie, the distance from the UAV to Bob, $d_b$, first decreases and then increases. Following Lemma~\ref{lemma1}, we note that the size of closed interval for the unique optimal $d_w$ increases with $\theta_w$, i.e., $d_w = L$ for $\theta_w = 0$ and $d_w \in [0, \infty]$ for $\theta_w = \pi/2$. We also note that the values of $L\cos(\theta_w)$ and $L/\cos(\theta_w)$ are critical, which divide the range of $d_w$ into three subintervals. Explicitly, $f(d_w, \theta_w)$ monotonically increases with $d_w$ for $\theta_w \in [0, L\cos(\theta_w)]$, and 
 monotonically decreases with $d_w$ for $\theta_w \in [L/\cos(\theta_w), \infty]$. The optimal $d_w$ that maximizes $f(d_w, \theta_w)$ is therefore within $[L\cos(\theta_w), L/\cos(\theta_w)]$. In addition, following the uniqueness of the optimal $d_w$, we can conclude that $f(d_w, \theta_w)$ monotonically increases with $d_w$ when $d_w \leq d_w^\circ(\theta_w)$ and $f(d_w, \theta_w)$ monotonically decreases with $d_w$ when $d_w^\circ(\theta_w) \leq d_w$.

We now tackle the optimal $\theta_w$ that maximizes $\gamma_b(P, d_w, \theta_w)$ for a given $d_w$ without any constraint. Noting $\gamma_b(P, d_w, \theta_w) = Pf(d_w, \theta_w)/\sigma_b^2$ again, the corresponding optimization problem is given by
\begin{align}\label{opt_thetaw_nocovert}
 \argmax_{\theta_w} ~~&f(d_w, \theta_w).
\end{align}
We have the following lemma detailing some properties of the solution to the optimization problem \eqref{opt_thetaw_nocovert}.

\begin{lemma}\label{lemma2}
When $d_w \leq L$ holds, the solution to the optimization problem \eqref{opt_thetaw_nocovert}, i.e., the optimal value of $\theta_w$ (denoted as $\theta_w^\circ(d_w)$) that maximizes $f(d_w, \theta_w)$ for a given $d_w$, is unique and is the unique solution in $\theta_w$ to the equation \eqref{optimal_thetaw_solution}
\begin{align} \label{optimal_thetaw_solution}
\xi_{{\textrm{L}}} L \sin(\theta_w)  \!+\!\frac{180 b}{\pi}\frac{\mathcal{G}(\theta_w)}{|L\!-\! d_w \cos(\theta_w)|}\left[1\!-\! p_b(d_w, \theta_w)\right] = 0,
\end{align}
where $\mathcal{G}(\theta_w) $ given by
\begin{align}\label{G_thetaw_de}
\mathcal{G}(\theta_w) = (L^2 + d_w^2)\cos(\theta_w) - d_w L \left[1+\cos^2(\theta_w)\right].
\end{align}
The optimal value, $\theta_w^\circ(d_w)$, lies in the closed interval $[0, \arccos (d_w/L)]$, and ${\partial f(d_w, \theta_w)}/{\partial \theta_w} < 0$ for $\arccos(d_w/L) \leq \theta_w$.
\end{lemma}
\begin{IEEEproof}
The detailed proof is presented in Appendix~B.
\end{IEEEproof}

 Intuitively, the results in Lemma~\ref{lemma2} are due to the fact that, for a fixed $d_w$ with $d_w \leq L$, as the angle $\theta_w$ increases (i.e., the UAV flies higher while keeping the same distance to Willie), the distance from the UAV to Bob, $d_b$, monotonically increases, while the angle from the UAV to Bob, $\theta_b$, first increases and then decreases with $\theta_w = \arccos (d_w/L)$ as the turning value. This is also the reason why the objective function monotonically decreases with $\theta_w$ for $\arccos (d_w/L) < \theta_w$. We note that the value of $\arccos (d_w/L)$ decreases with $d_w$, which means that the region where the UAV's optimal location belongs shrinks towards to Bob as $d_w$ increases. We can conclude that the optimal location of the UAV is at Bob's location for $d_w = L$.

\begin{lemma}\label{lemma3}
When $d_w>L$ holds, the solution to the optimization problem \eqref{opt_thetaw_nocovert}, i.e., the optimal value of $\theta_w$ (denoted as $\theta_w^\circ(d_w)$) that maximizes $f(d_w, \theta_w)$ for a given $d_w$, is unique and is the unique solution in $\theta_w$ to the equation \eqref{optimal_thetaw_solution2}
\begin{align} \label{optimal_thetaw_solution2}
\xi_{{\textrm{L}}} L \sin(\theta_w)  \!+\!\frac{180 b}{\pi}\frac{\mathcal{G}(\theta_w)}{|L\!-\! d_w \cos(\theta_w)|}\left[1\!-\! p_b(d_w, \theta_w)\right] = 0.
\end{align}
where $\mathcal{G}(\theta_w)$ is given by \eqref{G_thetaw_de}, $\theta_w^\circ(d_w)$ lies in the closed interval $[0, \arccos (L/d_w)]$, and the result ${\partial f(d_w,\theta_w)}/{\partial \theta_w} < 0$ holds for $\arccos(L/d_w) \leq \theta_w$.
\end{lemma}
\begin{IEEEproof}
The detailed proof is similar to that of Lemma~\ref{lemma2}, except that the angle $\arccos(L/d_w)$ is determined by solving $\mathcal{G}(\theta_w) = 0$ as per \eqref{G_thetaw_de} within the value range of $\cos(\theta_w)$.
\end{IEEEproof}

We note that the location $(d_w, \arccos(L/d_w))$ (in polar coordinates) is directly above Bob when $L < d_w$. Therefore, as per Lemma~\ref{lemma3} we know that the UAV's optimal location is on the left side of Bob (while Willie is on the right side of Bob) as shown in Fig.~\ref{fig:fig1}. We next solve the original optimization problem \eqref{opt_2D} based on the results presented in Lemma~\ref{lemma1}, Lemma~\ref{lemma2}, and Lemma~\ref{lemma3}. We present the solution in six different scenarios with different relationships among $\overline{d}_w$, $\underline{d}_w$, $\underline{\theta}_w$, and $L$. Specifically, we gradually reduce the value of $L$ (i.e., the distance between Bob and Willie) as we go from Scenario 1 to Scenario 6.

\subsection{Scenario 1 with $\underline{d}_w < \overline{d}_w \leq  L\cos(\underline{\theta}_w) <L/\cos(\underline{\theta}_w)$}

\begin{figure}[!t]
    \begin{center}
        \includegraphics[width=3.5in]{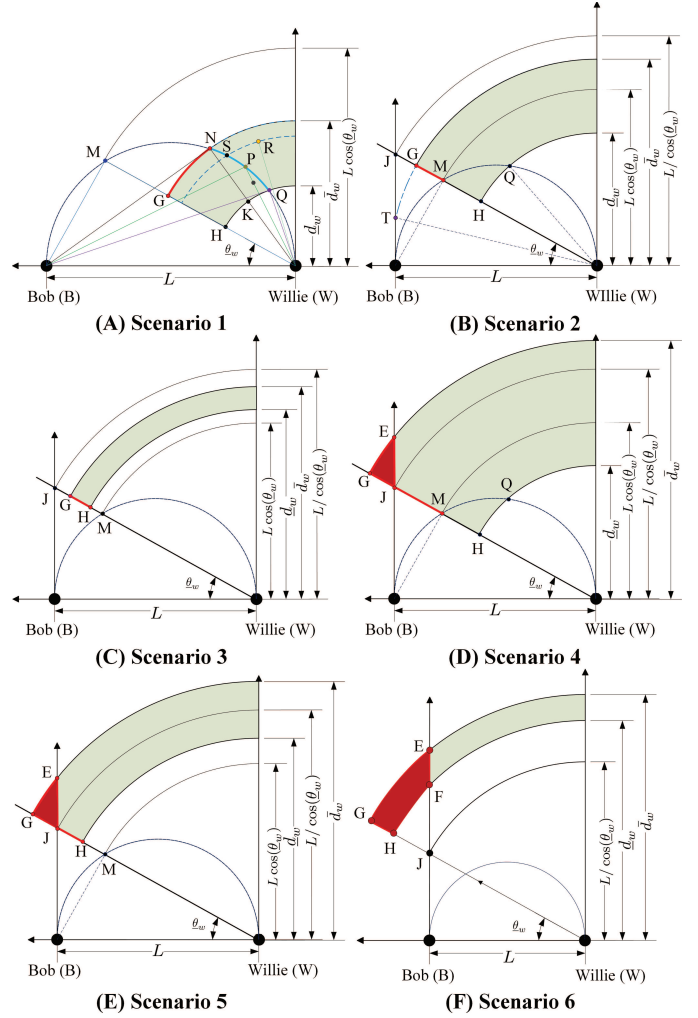}
        \caption{The optimal location of the UAV is on the red line segment or within the region closed by the red line segments in six different scenarios.}
        \label{fig:figs6scenarios}
    \end{center}
\end{figure}

This Scenario is depicted in Fig.~\ref{fig:figs6scenarios}(A) in which the distance between Bob and Willie, $L$, is much larger than $\bar{d}_w$. This is the most common scenario in practice, where Bob is a long way from the surveillance area.  

As shown in Fig.~\ref{fig:figs6scenarios}(A), the light green region is the 
feasible set for the UAV's location, which is determined by the constraints $\underline{d}_w \leq d_w \leq \overline{d}_w$ and $\underline{\theta}_w \leq \theta_w \leq \frac{\pi}{2}$. The semi-circle with $L$ as its diameter is determined by the locations of Bob and Willie. The points M, N and Q all lie on this semi-circle and are with different distances from Willie, which are $L\cos(\underline{\theta}_w), \overline{d}_w,$ and $\underline{d}_w,$ respectively. Further, the angles $\angle BMW = \angle BNW = \angle BQW = 90^\circ$, $\angle BWN = \arccos(\overline{d}_w/L)$ and $\angle BWQ = \arccos(\underline{d}_w/L)$.

Following \eqref{opt_2D}, the optimization problem in the specific scenario as shown in Fig.~\ref{fig:figs6scenarios}(A) is
\begin{subequations}\label{opt_2L_case1}
\begin{align}
 \argmax_{P, d_w, \theta_w} ~~&\gamma_b(P, d_w, \theta_w),\\
~~~~\text{s.t.} ~~& \mathcal{D}_{01} \leq 2 \epsilon^2,\\
&P \leq P_m,\\
&\underline{d}_w \leq d_w \leq \overline{d}_w, \label{distance_con2_case1} \\
&\underline{\theta}_w \leq \theta_w \leq \frac{\pi}{2}.\label{angle_con2_case1}
\end{align}
\end{subequations}

In order to solve the above optimization problem, we first determine a feasible set of a significantly reduced size for the UAV's optimal location in the following theorem.
\begin{theorem}\label{theorem_caseA}
The UAV's optimal location $(d_w^\ast, \theta_w^\ast)$ is on the arc GN shown in Fig.~\ref{fig:figs6scenarios}(A), i.e., $d_w^\ast = \overline{d}_w$ and $\theta_w^\ast$ satisfies $\underline{\theta}_w \leq \theta_w^\ast\leq \arccos(\frac{\overline{d}_w}{L})$, regardless of the maximum transmit power constraint $P \leq P_m$.
\end{theorem}

\begin{IEEEproof}
We prove this theorem by first restricting the feasible set of $(d_w,\theta_w)$ to the subset ${\cal F}$ defined by
\begin{align}
  {\cal F} = \{(d_w, \theta_w): \underline{d}_w \leq d_w \leq \overline{d}_w, \underline{\theta}_w \leq \theta_w < \arccos(\frac{\overline{d}_w}{L})\},\notag
\end{align}
which is the green region to the left side of the straight line segment NK in Fig.~\ref{fig:figs6scenarios}(A). Over this subregion, we show that the optimal point must lie on the red arc GN. We then show that the optimal solution cannot lie outside of ${\cal F}$, {i.e.,} it cannot lie in the green region on the right side of the line segment NK shown in Fig.~\ref{fig:figs6scenarios}(A).

For $(d_w, \theta_w) \in {\cal F}$, we have $L\cos(\theta_w) > d_w$, due to the following inequalities holding:
\begin{align*}
    d_w \leq \overline{d}_w < L \cos\left(\underline{\theta}_w\right) < L \cos(\theta_w).
\end{align*}
The second inequality holds because we are in Scenario 1.
Lemma~\ref{lemma1} shows that the objective function $\gamma_b(P, d_w, \theta_w)$ monotonically increases with $d_w$, for fixed $\theta_w$, when $L\cos(\theta_w) > d_w$. We note that $\mathcal{D}_{01}$ in the constraint \eqref{covertness_con2} is a monotonically decreasing function of $d_w$, since the received power $\bar{P} = P d_w^{\xi_{{\textrm{L}}}}p_w + P d_w^{\xi_{{\textrm{N}}}}$ (which appears in $\mathcal{D}_{01}$) decreases with $d_w$ since  $\xi_{{\textrm{L}}} <0$ and $\xi_{{\textrm{N}}} <0$. Considering the constraint $\underline{d}_w \leq d_w \leq \overline{d}_w$, we can conclude that $\overline{d}_w$ is the optimal $d_w$ for any given $\theta_w$ in this region, regardless of the maximum transmit power constraint, which means that the optimal solution within ${\cal F}$ lies on the red arc GN.

We now show that the optimal solution cannot lie outside the subset ${\cal F}$. Note that if $(d_w,\theta_w)$ lies in the feasible region, but outside the set ${\cal F}$, then $\theta_w > \arccos(\overline{d}_w/L)$.

First, suppose that $(d_w,\theta_w)$ lies above the line segment NK but below the blue arc NQ. The point $(L \cos(\theta_w), \theta_w)$ has the same angle $\theta$, but a greater distance from Willie, such that it lies on the blue arc (illustrated by point P in the Figure). By Lemma 1, the objective function is higher at P than at the considered point $(d_w,\theta_w)$. Since $\mathcal{D}_{01}$ in the constraint \eqref{covertness_con2} is a monotonically decreasing function of $d_w$, it follows that $(d_w,\theta_w)$ cannot be the optimal solution.

Next, consider a point $(d_w, \theta_w)$ that lies above the blue arc NQ, as illustrated by point R in Fig.~\ref{fig:figs6scenarios}(A). The point S has the same distance, $d_w$, from Willie as R, but a smaller angle, $\arccos(d_w/L)$, which means it lies on the blue arc NQ. Lemma~\ref{lemma2} shows that $f(d_w, \theta_w)$ monotonically decreases with $\theta_w$ for a fixed $d_w$ when $\theta_w > \arccos(d_w/L)$. Since $\mathcal{D}_{01}$ monotonically increases with $\theta_w$, we can conclude that the the objective function value at the point S is higher than at point R, and the value of $\mathcal{D}_{01}$ is smaller. It follows that the optimal solution cannot lie strictly above the blue arc NQ.

Finally, consider any point $(d_w, \theta_w)$ that lies on the blue arc NQ but strictly to the right of point N, {i.e.} with $d_w < \overline{d_w}$. Lemma~\ref{lemma1} shows that such a point cannot be optimal, since the objective function increases with $d_w$ for fixed angle $\theta_w$, and $\mathcal{D}_{01}$ decreases.

We conclude that the optimal solution must lie on the red arc GN.
\end{IEEEproof}

Following Theorem~\ref{theorem_caseA}, the solution to the optimization problem \eqref{opt_2L_case1} is presented in the following proposition.
\begin{proposition}\label{proposition1}
The UAV's optimal location and transmit power that jointly maximize $\gamma_b(P, d_w, \theta_w)$ subject to the constraints \eqref{covertness_con2}, \eqref{power_max_con}, \eqref{distance_con2}, and \eqref{angle_con}, 
are given in Table~\ref{tab1}, where we recall that $\theta_w^\circ(\overline{d}_w)$ is the optimal $\theta_w$ that maximizes $\gamma_b(P, \overline{d}_w, \theta_w)$ (obtained in Lemma~\ref{lemma2}), $\mathcal{D}_{01}(P, d_w, \theta_w)$ is the KL divergence $\mathcal{D}_{01}$, between ${\mathbb P}_0$ and ${\mathbb P}_1$, as an explicit function of $P$, $d_w$, and $\theta_w$, and $P^{\epsilon}$ is the unique value of $P$ that ensures $\mathcal{D}_{01}(P, \overline{d}_w, \underline{\theta}_w) = 2 \epsilon^2$. The values $P^j$ and $\theta_w^j$ are the unique maximizers of the following optimization problem
\begin{subequations}\label{opt_2L_case1_sub}
\begin{align}
 \argmax_{P, \theta_w} ~~&\gamma_b(P, \overline{d}_w, \theta_w),\\
~~~~\text{s.t.} ~~& \mathcal{D}_{01}(P, \overline{d}_w, \theta_w) = 2 \epsilon^2, \label{covertness_con_equal}\\
&P \leq P_m, \label{power_max_con_sub}\\
&\max\{\underline{\theta}_w,\theta_w^{\epsilon}\} \leq \theta_w \leq \theta_w^\circ(\overline{d}_w),\label{angle_con_sub}
\end{align}
\end{subequations}
where $\theta_w^{\epsilon}$ is the unique solution in $\theta_w$ to the equation $\mathcal{D}_{01}(P_m, \overline{d}_w, \theta_w) = 2 \epsilon^2$.

\begin{table}[ht]
\caption{{Solutions to the optimization problem \eqref{opt_2L_case1}.}}
\label{tab1}%
\centering
\begin{tabular}{|l||c|} \hline
    \textbf{Solutions} ($d_w^\ast, \theta_w^\ast, P^\ast$) & \textbf{Conditions} \\ \hline \hline
    Case~A: ($\overline{d}_w, \underline{\theta}_w, P_m$) &  \makecell{$\theta_w^\circ(\overline{d}_w) < \underline{\theta}_w$~\\~$\mathcal{D}_{01}(P_m, \overline{d}_w, \underline{\theta}_w) \leq 2 \epsilon^2$}  \\ \hline
    Case~B: ($\overline{d}_w, \underline{\theta}_w, P^{\epsilon}$) &  \makecell{$\theta_w^\circ(\overline{d}_w) < \underline{\theta}_w$~\\~$\mathcal{D}_{01}(P_m, \overline{d}_w, \underline{\theta}_w) > 2 \epsilon^2$}  \\ \hline
    Case~C: ($\overline{d}_w, \theta_w^\circ(\overline{d}_w), P_m$) &  \makecell{$\underline{\theta}_w \leq \theta_w^\circ(\overline{d}_w) \leq \arccos(\frac{\overline{d}_w}{L})$~\\~$\mathcal{D}_{01}(P_m, \overline{d}_w, \theta_w^\circ(\overline{d}_w)) \leq 2 \epsilon^2$}  \\ \hline
    Case~D: ($\overline{d}_w, \theta_w^j, P^j$) &  \makecell{$\underline{\theta}_w \leq \theta_w^\circ(\overline{d}_w) \leq \arccos(\frac{\overline{d}_w}{L})$~\\~$\mathcal{D}_{01}(P_m, \overline{d}_w, \theta_w^\circ(\overline{d}_w)) > 2 \epsilon^2$}  \\ \hline
    \hline
\end{tabular}
\end{table}
\end{proposition}

\begin{IEEEproof}
When $\theta_w^\circ(\overline{d}_w) < \underline{\theta}_w$, we have $\theta_w^\ast = \underline{\theta}_w$, where we recall that $\theta_w^\circ(\overline{d}_w)$ is the optimal $\theta_w$ for $d_w = \overline{d}_w$ without any constraint. This is due to the fact that, with $\theta_w^\circ(\overline{d}_w) < \underline{\theta}_w$, both the objective function $\gamma_b(P, \overline{d}_w, \theta_w)$ and the KL divergence $\mathcal{D}_{01}(P, \overline{d}_w, \theta_w)$ monnotonically decrease with $\theta_w$ for $\underline{\theta}_w \leq \theta_w \leq \arccos(\frac{\overline{d}_w}{L})$. In this case, if $\mathcal{D}_{01}(P_m, \overline{d}_w, \underline{\theta}_w) \leq 2 \epsilon^2$, we have $P^\ast = P_m$. Otherwise, the optimal value of $P$ is $P^{\epsilon}$, which satisfies
$\mathcal{D}_{01}(P, \overline{d}_w, \underline{\theta}_w) = 2 \epsilon^2$.

We now focus on the case with $\underline{\theta}_w \leq \theta_w^\circ(\overline{d}_w) \leq \arccos(\frac{\overline{d}_w}{L})$. In this case, we have $\theta_w^\ast = \theta_w^\circ(\overline{d}_w)$ and $P^\ast = P_m$, if $\mathcal{D}_{01}(P_m, \overline{d}_w, \theta_w^\circ(\overline{d}_w)) \leq 2 \epsilon^2$, which is due to the fact that the covertness constraint is not active at this point. In this case, if  $\mathcal{D}_{01}(P_m, \overline{d}_w, \theta_w^\circ(\overline{d}_w)) > 2 \epsilon^2$, the optimal values of $\theta_w$ and $P$ are the ones that jointly maximize the objective function $\gamma_b(P, \overline{d}_w, \theta_w)$ subject to $\mathcal{D}_{01}(P, \overline{d}_w, \theta_w) \leq 2 \epsilon^2$, $\max\{\underline{\theta}_w,\theta_w^{\epsilon}\} \leq \theta_w \leq \theta_w^\circ(\overline{d}_w)$, and $0<P \leq P_m$. We note that the equality in $\mathcal{D}_{01}(P, \overline{d}_w, \theta_w) \leq 2 \epsilon^2$ can always be satisfied, since the objective function $\gamma_b(P, \overline{d}_w, \theta_w)$ and the KL divergence $\mathcal{D}_{01}(P, \overline{d}_w, \theta_w)$ are both monotonically decreasing functions of $\theta_w$, while they are monotonically increasing functions of the transmit power $P$. Specifically, the optimization problem is given in \eqref{opt_2L_case1_sub}, which can be efficiently solved by a 
1-dimensional search for $\theta_w$ over the interval $\underline{\theta}_w \leq \theta_w \leq \theta_w^\circ(\overline{d}_w)$ since $P$ is uniquely determined by \eqref{covertness_con_equal}.
\end{IEEEproof}

In the following subsections we present the feasible regions of significantly reduced size for the UAV's optimal location for the remaining five scenarios. The proofs 
follow along similar lines to the proof of Theorem~\ref{theorem_caseA} but are omitted due to space considerations.

\subsection{Scenario 2 with $\underline{d}_w  \leq  L\cos(\underline{\theta}_w) \leq \overline{d}_w \leq L/\cos(\underline{\theta}_w)$}

As shown in Fig.~\ref{fig:figs6scenarios}(B), Scenario 2 with $\underline{d}_w  \leq  L\cos(\underline{\theta}_w) \leq \overline{d}_w \leq L/\cos(\underline{\theta}_w)$ is obtained by decreasing the distance, $L$, between Bob and Willie from that considered in Scenario 1. In Scenario 2, we can conclude that the optimal location of the UAV is on the line segment GM (shown in red) in Fig.~\ref{fig:figs6scenarios}(B), due to the following reasons.
\begin{itemize}
\item  In the green shaded region with $\underline{d}_w \leq d_w \leq L\cos(\underline{\theta}_w)$ and $\underline{\theta}_w \leq \theta_w \leq \pi/2$, the best location of the UAV is on the point M, which follows from facts detailed 
in the proof of Theorem~\ref{theorem_caseA}.
\item In the region with $L\cos(\underline{\theta}_w) \leq d_w \leq \overline{d}_w$, the best location the UAV is on the line segment GM, 
which follows from results presented in Lemma~\ref{lemma2} and Lemma~\ref{lemma3}.
\end{itemize}

\subsection{Scenario 3 with $L\cos(\underline{\theta}_w) \leq \underline{d}_w  \leq  \overline{d}_w \leq L/\cos(\underline{\theta}_w)$}

Following Scenario~2, if we further decrease the distance, $L$, between Bob and Willie, we may have $L\cos(\underline{\theta}_w) \leq \underline{d}_w  \leq  \overline{d}_w \leq L/\cos(\underline{\theta}_w)$ which is Scenario~3 as shown in Fig.~\ref{fig:figs6scenarios}(C). Alternativey, we may have $\underline{d}_w \leq L\cos(\underline{\theta}_w) < L/\cos(\underline{\theta}_w)  \leq \overline{d}_w$ which is Scenario~4 in Fig.~\ref{fig:figs6scenarios}(D).

In Scenario 3, for any possible location within the green region on the right side of the line segment GH, we can find one location on the line segment GH that achieves a higher objective function $\gamma_b(P, d_w, \theta_w)$. This is due to the following two facts.
\begin{itemize}
\item  For any given fixed $d_w$ within this area (i.e., the green region in Fig.~\ref{fig:figs6scenarios}(C)), the distance to Bob, $d_b$, increases with $\theta_w$, while the angle to Bob, $\theta_b$, monotonically decreases with $\theta_w$ (due to that there is no intersection between the green region and the semi-circle), which leads to that the objective function $\gamma_b(P, d_w, \theta_w)$ monotonically decreases with $\theta_w$ for any fixed $d_w$ in the green region.
\item The KL divergence $\mathcal{D}_{01}$ is a monotonically increasing function of $\theta_w$ for any fixed $d_w$ in the green region, since the probability $p_w$ to have the LoS component in the channel from the UAV to Willie monotonically increases with $\theta_w$.
\end{itemize}

Therefore, we can conclude that the optimal location of the UAV is on the line segment GH in this scenario. We also note that, if the covertness constraint can be satisfied with $P = P_m$ at the point H, we can conclude that the optimal location is on the point H. This is due to the fact that, for a fixed transmit power, both the objective function $\gamma_b(P, d_w, \theta_w)$ and the KL divergence  $\mathcal{D}_{01}$ in the covertness constraint $\mathcal{D}_{01} \leq 2 \epsilon^2$ monotonically decrease with $d_w$ on the line segment GH, as shown in Lemma~\ref{lemma1}.

\subsection{Scenario 4 with $\underline{d}_w \leq L\cos(\underline{\theta}_w) < L/\cos(\underline{\theta}_w)  \leq \overline{d}_w$}

Scenario~4 can arise instead of Scenario 3, from decreasing $L$ from Scenario 2. Scenario~4 occurs if $\underline{d}_w \leq L\cos(\underline{\theta}_w) < L/\cos(\underline{\theta}_w)  \leq \overline{d}_w$ when decreasing $L$ from Scenario 2. Scenario~4 is illustrated in Fig.~\ref{fig:figs6scenarios}(D). In this scenario, we can conclude that the optimal location of the UAV is within the region EGJ or on the line segment JM (shown in red), due to  Lemma~\ref{lemma3} and Theorem~\ref{theorem_caseA}.

\subsection{Scenario 5 with $L\cos(\underline{\theta}_w) \leq \underline{d}_w  \leq L/\cos(\underline{\theta}_w)  \leq \overline{d}_w$}

If we further decrease the distance, $L$, between Bob and Willie, beyond Scenario 3 or 4, we will have $L\cos(\underline{\theta}_w) \leq \underline{d}_w \leq L/\cos(\underline{\theta}_w)  \leq \overline{d}_w$, which is Scenario~5  as shown in Fig.~\ref{fig:figs6scenarios}(E). In this scenario, we can conclude that the optimal location of the UAV is within the region EGJ or on the line segment JH (shown in red), due to similar reasons for achieving the results in Scenario~4.

\subsection{Scenario 6 with $L\cos(\underline{\theta}_w) < L/\cos(\underline{\theta}_w)  \leq \underline{d}_w < \overline{d}_w$}

Following Scenario~5, if we continue to reduce $L$, we will obtain Scenario~6 with $L\cos(\underline{\theta}_w) < L/\cos(\underline{\theta}_w)  \leq \underline{d}_w < \overline{d}_w$ as shown in Fig.~\ref{fig:figs6scenarios}(F). In this scenario, following Lemma~\ref{lemma3} and the discussions on Scenario~5, we can conclude that the UAV's optimal location is on the left side of Bob, i.e., within the closed region GEFH.

\subsection{Heuristic Approaches}\label{sec-heuristic}

From a practical point of view, two simple alternatives to solving \eqref{opt_2D} present themselves. The first is to simply fly the UAV at the closest point to Bob within the feasible region, regardless of the scenario. The second is to choose the point in the feasible region with the largest angle to Bob, since this maximizes the probability of a LoS to Bob. In Section ~\ref{section_num}, we compare these heuristic suboptimal approaches with our approach, and show that in some cases they are able to achieve close to optimal, but that in other cases significant gains can be achieved by employing our optimal approaches.

\subsection{Summary of Results}

The following general conclusions provide a summary of the results from the six scenarios.
\begin{itemize}
  \item As long as the whole feasible region of the optimal location is on the upper right side of Bob, the 2-dimensional location optimization problem reduces to a 1-dimensional optimization problem, which can be seen from Scenarios 1, 2, and 3 as shown in Fig.~\ref{fig:figs6scenarios}.
  \item We note that Scenario~1 is the most common scenario, where Bob is far from Willie and thus the wireless communication from the UAV to Bob is desirable for conveying urgent information about the surveillance outcomes. However, in this work we consider all the scenarios to present a whole picture about the UAV's location optimization for covert wireless communications.
 \end{itemize}

\section{Optimal Height and Transmit Power for A Vertical UAV}\label{section_Vertical}

In this section, we focus on a special scenario with a vertical UAV, i.e., the UAV can only adjust its flying height. This scenario arises if the surveillance equipment (e.g., a camera) can only undertake its surveillance tasks when the UAV is directly above the target. We note that this 
is a special case of Scenario 3 discussed in the previous section by setting $\underline{\theta}_w = 90^\circ.$ Considering this vertical UAV, in this section we determine the optimal height, $h$, and transmit power, $P$. We denote the lower and upper bounds on the height by $\underline{h}$, and $\overline{h}$, respectively.  To simplify the presentation, we denote the effective SNR (denoted as $\gamma_b(P, d_w, \theta_w)$ in the previous section) as $\gamma_b(P, h)$ in this section.

\subsection{Optimal Height without Any Constraint}

In order to determine the optimal height and transmit power of the UAV to conduct covert transmission, similar to Section II-A, in this subsection we derive the optimal height of the UAV for transmitting to Bob without any constraint. We note that there is an optimal height for the UAV Alice, which maximizes the effective SNR $\gamma_b(P, h)$ at Bob, since both $d_b$ and $p_b$ monotonically increase with $h$. To facilitate determining the optimal height of the UAV with a covertness constraint, we first identify the optimal height $h$ that maximizes $\gamma_b(P, h)$ for a given transmit power $P$. The corresponding optimization problem is given by
\begin{align}\label{opt_height_nocovert}
 \argmax_{h} ~~&\gamma_b(P, h),
\end{align}
where we recall that $\gamma_b(P, h) = P d_b^{\xi_{{\textrm{L}}}} p_b/{\sigma_b^2}$ and $p_b$ is given in (2). Following Lemma~\ref{lemma1}, the solution to the optimization problem \eqref{opt_height_nocovert} is given in the following lemma.

\begin{lemma}\label{lemma4}
The optimal $h$, which we denote by $h^{\dag}$, that maximizes $\gamma_b(P, h)$ is the unique solution to the following fixed-point equation:
\begin{align}
h\!=\! \!-\!\frac{180 b L}{\pi \xi_{{\textrm{L}}}}\left[1\!-\!\frac{1}{1 \!+\! a \exp(-b[\frac{180}{\pi}\tan^{-1}(\frac{h}{L}) \!-\! a])}\right].
\end{align}
\end{lemma}
\begin{IEEEproof} Obtained from  Lemma~\ref{lemma1} by setting $\theta_w = 90^\circ$.
\end{IEEEproof}

From Lemma~\ref{lemma4}, we note that the optimal height $h^{\dag}$ of the UAV without any constraint is independent of the UAV's transmit power $P$ and the AWGN variance at Bob (i.e., $\sigma_b^2$), and is solely determined by the environment parameters (e.g., $\xi_{{\textrm{L}}}$, $a$, $b$, and $L$).

\subsection{Optimal Height with Covertness, Transmit Power, and Surveillance Constraints}

The optimization problem at the UAV is given by
\begin{subequations}\label{opt_vertical}
\begin{align}
 \argmax_{P, h} ~~&\gamma_b(P, h),\\
~~~~\text{s.t.} ~~& \mathcal{D}_{01} \leq 2 \epsilon^2,\\
&P \leq P_m,\\
&h \leq \overline{h}, \label{surveillance_con} \\
&h \geq \underline{h}, \label{distance_con}
\end{align}
\end{subequations}
where we recall that $\overline{h}$ is the maximum value of the UAV's heigh, and $\underline{h}$ is the minimum value. 

In order to facilitate presenting the solution to the optimization problem \eqref{opt_vertical}, we first write $\mathcal{D}_{01}$ as an explicit function of $h$ and $P$, which is $\mathcal{D}_{01}(P, h)$. Following \eqref{D01_ex}, we note that, for a fixed $P$ and $p_w = 1$ due to $\theta_w = 90^\circ$, $\mathcal{D}_{01}(P, h)$ is a monotonically decreasing function of $h$, since, for a fixed transmit power $P$, the received power $\bar{P} = P d_w^{\xi_{{\textrm{L}}}}p_w + P d_w^{\xi_{{\textrm{N}}}}$ in \eqref{D01_ex} monotonically decreases with $h$. We denote the unique solution of $h$ to $\mathcal{D}_{01}(P_m, h) = 2 \epsilon^2$ as $h^{\ddag}$, where we recall that $P_m$ is the maximum transmit power of the UAV. Likewise, $\mathcal{D}_{01}(P, h)$ is a monotonically increasing function of $P$ for a fixed $h$. As such, we denote the unique solution of $P$ to $\mathcal{D}_{01}(P, \overline{h}) = 2 \epsilon^2$ as $P^{\dag}$, where we recall that $\overline{h}$ is the upper bound on $h$ (i.e., $h\leq \overline{h}$). We note that $h^{\ddag}$ and $P^{\dag}$ can be approximately achieved in closed-form expressions by using a similar approach presented in~\cite{shu2019delay}. We next present the solution to the optimization problem \eqref{opt_vertical} in the following theorem.

\begin{theorem}
The optimal height $h^\ast$ and the optimal transmit power $P^\ast$ that jointly maximize $\gamma_b(P, h)$ subject to the constraints \eqref{covertness_con2}, \eqref{power_max_con}, \eqref{surveillance_con}, and \eqref{distance_con} are given in Table II as below.
\begin{table}[ht]
\label{tab:time}%
\caption{{Solutions to the optimization problem  \eqref{opt_vertical}.}}
\centering
\begin{tabular}{|l||c|} \hline
    \textbf{Solutions} ($h^\ast, P^\ast$) & \textbf{Conditions} \\ \hline \hline
    Case~1: ($\underline{h}, P_m$) &  $h^\dag \leq \underline{h}$~and~$h^\ddag \leq \underline{h}$  \\ \hline
    Case~2: ($h^\ddag, P_m$) &  \tabincell{c}{$\underline{h}<h^\ddag < \overline{h}$ and $h^\dag<h^\ddag$} \\\hline
    Case~3: ($\overline{h}, P^\dag$) &  \tabincell{c}{$h^\ddag \geq \overline{h}$} \\\hline
    Case~4: ($h^\dag, P_m$) &  \tabincell{c}{$\underline{h}<h^\dag < \overline{h}$~and~$h^\ddag<h^\dag$} \\\hline
    Case~5: ($\overline{h}, P_m$) &  \tabincell{c}{$h^\dag \geq \overline{h}$~and~$h^\ddag \leq \overline{h}$} \\\hline
    \hline
\end{tabular}
\end{table}
\end{theorem}
\begin{IEEEproof}
The detailed proof can be found in our conference version of this work \cite{yan2019hiding}.
\end{IEEEproof}

Following Theorem~2, we note that the UAV's optimal height with covertness constraint $h^\ast$ is indeed a function of its transmit power, although its optimal height without any constraint is independent of its transmit power as shown in Lemma~\ref{lemma4}. We also note that, when the maximum transmit power of the UAV is sufficiently large, we should have the solution in Case~3, since $h^\ddag$ is the solution to $\mathcal{D}_{01}(P_m, h) = 2 \epsilon^2$ and it monotonically increases with $P_m$ as per \eqref{D01_ex}. As expected, from Theorem~2 we can see that $h^\ast$ is a function of $h^\ddag$, which leads to the fact that $h^\ast$ highly depends on the covertness constraint (i.e., the value of $\epsilon$).

\section{Numerical Results}\label{section_num}

\begin{figure}[!t]
    \begin{center}
        \includegraphics[width=3.4in, height=2.7in]{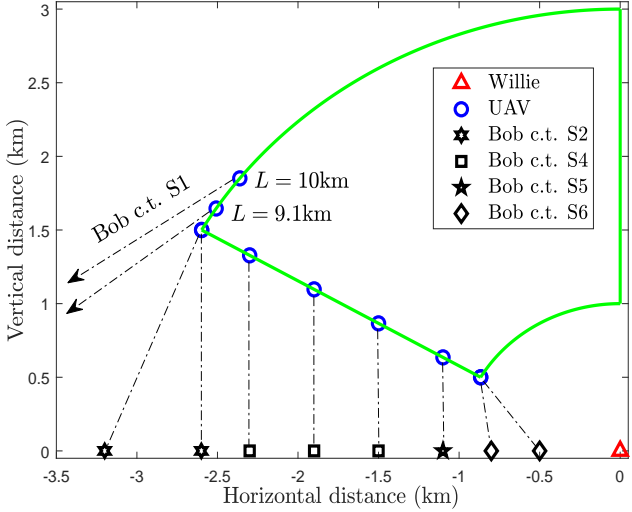}
        \caption{The optimal location of the UAV for different values of the distance, $L$, between Bob and Willie, corresponding to (c.t.) different scenarios, where $a = 4.88$, $b = 0.429$, $\xi_L = -2$, $\xi_N = -3$, $P_m = 20$dB, $\epsilon = 0.01$, $\sigma_b^2  = -90$dB, $ \sigma_w^2 = -60$dB, $n = 200$, $\underline{\theta}_w=\pi/6$, $\overline{d}_w=3$km, and $\underline{d}_w = 1$km.}
        \label{fig:fig5}
    \end{center}
\end{figure}

In Fig.~\ref{fig:fig5}, we plot the UAV's optimal location for different values of $L$, the distance between Bob and Willie, where the region bounded by the solid green lines is the feasible region for the UAV's location. We use a dashed line to connect each of Bob's locations to the corresponding UAV's optimal location. If we consider the locations in sequence, as we decrease $L$ from $10$ km to $0.5$ km, we observe that the optimal point moves from a location that is the maximum distance from Willie to the minimum distance. At the first two points (when $L=10$ km, and $L=9.5$ km, respectively) we are in Scenario 1, and the maximum distance constraint is tight as is always the case in Scenario 1. The lower bound on the angle is not tight in these two cases, due to the fact that the probability of a LoS component to Bob is higher as the UAV flies higher. At the next point (on the corner, which corresponds to Scenario 2 in this case) the lower bound on the angle becomes tight. It remains tight for all the remaining points shown (which correspond to Scenarios 4, 5, and 6). At the last point (when $L=0.5$ km, which is in Scenario 6 in this case) the minimum distance constraint has become tight. 
We also observe that the UAV's optimal location is always on the border of the feasible region under the specific simulation settings of this figure. 
Note that in one case (in Scenario 6) the optimal solution is to fly the UAV at the nearest feasible location to Bob, and for others (Scenarios 4 and 5) the optimal location is almost directly above Bob (which is the closest point in the feasible region with the highest angle to Bob). 

\begin{figure}[!t]
    \begin{center}
        \includegraphics[width=3.4in, height=2.7in]{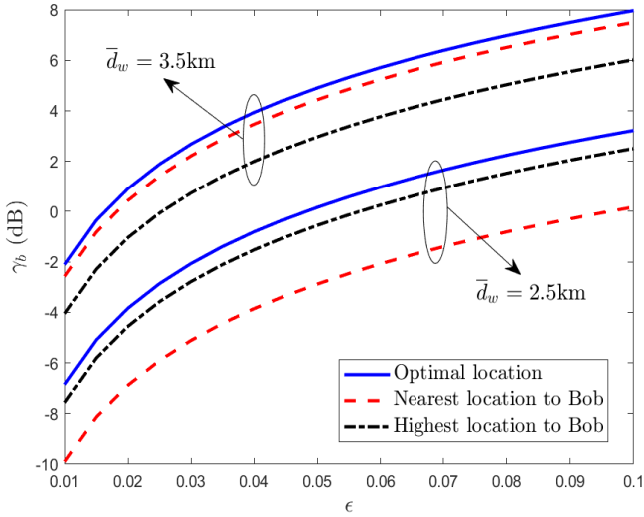}
        \caption{SNR, $\gamma_b$, versus $\epsilon$, for $L = 10$km,
         $P_m = 10$dBm, $\sigma_b^2  = -90$dBm, $ \sigma_w^2 = -60$dBm, $n = 200$, $\underline{\theta}_w=\pi/8$, $\overline{d}_w=3$km, and $\underline{d}_w = 1$km, and with channel parameters $a = 4.88$, $b = 0.429$, $\xi_L = -2$, $\xi_N = -3$.}
        \label{fig:fig_s1compare}
    \end{center}
\end{figure}

Fig.~\ref{fig:fig_s1compare} considers Scenario 1, and shows the achieved optimal values of Bob's effective SNR $\gamma_b$ 
when the UAV is in the optimal location. We observe that the optimal $\gamma_b$ monotonically increases with $\epsilon$. This is due to the fact that the maximum transmit power constraint is not active and the optimal transmit power increases with $\epsilon$ (as the covertness constraint becomes less onerous). We also plot the values of Bob's effective SNR for the two heuristic approaches described in Section~\ref{sec-heuristic}. We observe that neither heuristic scheme is optimal in general. When the distance to Willie is upper bounded by $3.5$ km, the nearest location to Bob is the better of the two. When the upper bound is $2.5$ km, the location with the highest angle to Bob is the better one. Additional numerical results, not reported here, suggest a third heuristic approach
would be to calculate both heuristic approaches and pick the best, although this is still not optimal.

\begin{figure}[!t]
    \begin{center}
        \includegraphics[width=3.4in, height=2.7in]{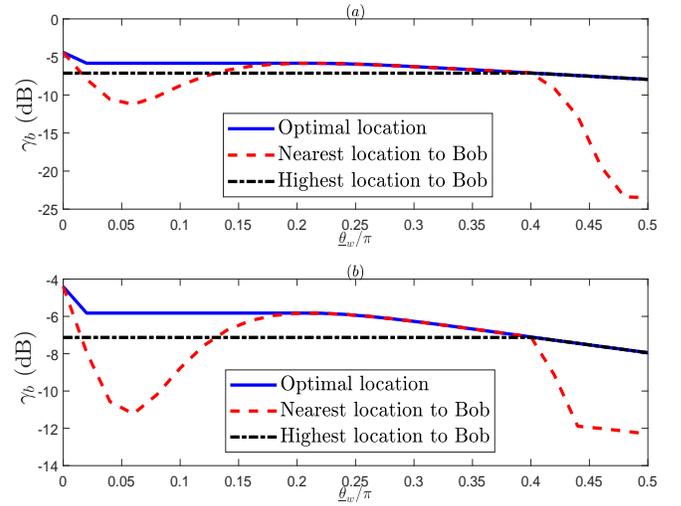}
        \caption{SNR, $\gamma_b$, versus $\epsilon$ for (a) $L = 10$km and (b) $L = 15$km.  With $P_m = 10$dBm, $\epsilon = 0.1$, $\sigma_b^2  = -90$dBm, $ \sigma_w^2 = -60$dBm, $n = 200$, $\underline{d}_w = 1$km, $\overline{d}_w=3$km, and with channel parameters $a = 4.88$, $b = 0.429$, $\xi_L = -2$, $\xi_N = -3$.}
        \label{fig:fig_s3dimention}
    \end{center}
\end{figure}

Fig.~\ref{fig:fig_s3dimention} plots the optimal SNR as a function of the lower angle constraint $\underline{\theta}_w$, for two different values of $L$. Note that $\underline{\theta}_w/\pi = 0.5$ corresponds to the vertical UAV scenario where there is only one degree of freedom: the UAV's height. We observe that the optimal value of $\gamma_b$ monotonically decreases with $\underline{\theta}_w$. The figure also shows results for the two heuristic schemes described in Section~\ref{sec-heuristic}. We observe that neither heuristic scheme is optimal across the whole range of $\underline{\theta}_w$.


\begin{figure}[!t]
    \begin{center}
        \includegraphics[width=3.4in, height=2.7in]{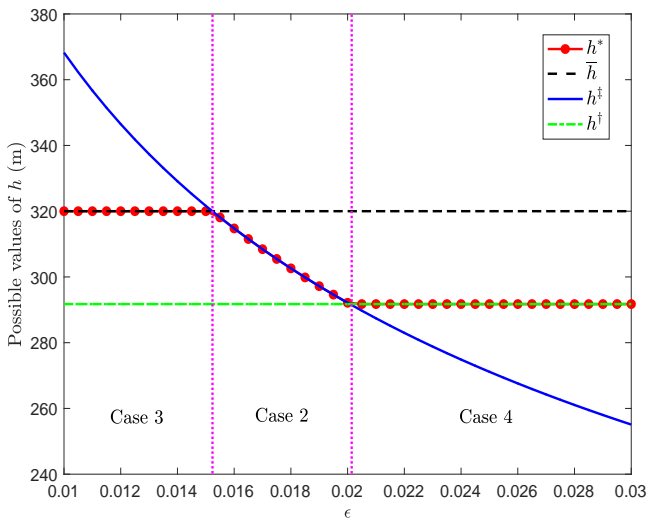}
        \caption{Possible values of UAV's height versus the covertness parameter $\epsilon$, where the S-curve parameters are $a = 4.88$ and $b = 0.429$, the path loss exponent is $\xi = -3$, the UAV's maximum transmit power is $P_m = 10$dBm, the AWGN power is $\sigma_w^2 = -40$dBm, the distance from Bob to Willie is $L = 1000$m, the number of the channel uses is $n = 200$, and the lower bound on the UAV's height is $\underline{h} = 100$m.}
        \label{fig:fig2}
    \end{center}
\end{figure}

In Fig.~\ref{fig:fig2}, we consider the vertical UAV scenario. Specifically, we plot possible values of the UAV's height $h$ versus the covertness parameter $\epsilon$, where we recall that a smaller $\epsilon$ represents a stricter covertness constraint. As expected, in this figure we first observe that neither $h^\dag$ nor $\overline{h}$ is a function of $\epsilon$, while $h^\ddag$ is indeed a function of $\epsilon$. We also observe that $h^\ddag$ monotonically decreases with $\epsilon$, which leads to the fact that the solutions to the optimal $h$ and $P$ move from Case~3 to Case~2 and then to Case~4 as $\epsilon$ increases. We note that in Case~3 and Case~4 we have $\mathcal{D}_{01} < 2 \epsilon^2$, while in Case~2 we have $\mathcal{D}_{01} = 2 \epsilon^2$. This is the reason why in Case~3 and Case~4 the optimal height $h^\ast$ does not directly depend on $\epsilon$, while in Case~2 we have that $h^\ast$ monotonically decreases with $\epsilon$. In general, this figure shows that the UAV's optimal height with covertness constraint decreases with $\epsilon$.

\section{Conclusion}\label{conclusion}

We have shown that covertness constraints on UAV communications in surveillance applications leads to six possible scenarios, with corresponding feasibility regions for the flying location of the UAV. We showed that these regions can be searched efficiently to find the optimal flying height and ground distance between the legitimate ground station and a potential eves dropper. We discussed two heuristic approaches to UAV placement, and showed that in some cases they are able to achieve close to optimal, but that in other cases significant gains can be achieved by employing our new search techniques. 

\section*{Acknowledgement}\label{ack}

We thank Prof. Dennis Goeckel for his contribution to the vertical UAV case in Section \ref{section_Vertical}.

\section*{Appendix A: Proof of Lemma~\ref{lemma1}}
\begin{IEEEproof}
Following \eqref{f_dw_thetaw}, the first derivative of $f(d_w, \theta_w)$ with respect to $d_w$ is derived as
\begin{align}
&\frac{\partial f(d_w, \theta_w)}{\partial d_w}
=\frac{p_b(d_w, \theta_w)\xi_{{\textrm{L}}}\left[d_w \!\!-\!\! L\cos(\theta_w)\right]}{\left[L^2 \!+\! d_w^2 \!\!-\!\! 2d_w L \cos(\theta_w)\right]^{1-\frac{\xi_{{\textrm{L}}}}{2}}}\!+\!p_b^2(d_w, \theta_w)\notag\\
&\left[L^2 \!+\! d_w^2 \!-\! 2d_w L \cos(\theta_w)\right]^{\frac{\xi_{{\textrm{L}}}}{2}}\frac{180}{\pi}ab \notag\\
&\times\exp\left\{\!-\!b \left[\frac{180}{\pi}\sin^{\!-\!1}\left(\frac{d_w \sin(\theta_w)}{\sqrt{L^2 \!+\! d_w^2 \!-\! 2d_w L \cos(\theta_w)}}\right)\right]\!-\!a\right\}\notag\\
&\times \frac{ \sin(\theta_w)}{|L- d_w \cos(\theta_w)|}\times \frac{L[L- d_w \cos(\theta_w)]}{L^2 + d_w^2 - 2d_w L \cos(\theta_w)},\\
&=p_b(d_w, \theta_w)[L^2 + d_w^2 - 2d_w L \cos(\theta_w)]^{\frac{\xi_{{\textrm{L}}}}{2}-1}\bar{f}(d_w),\label{f_dw_simple}
\end{align}
where $p_b(d_w, \theta_w)$ is the probability to have the LoS component in the channel from the UAV to Bob, given by
\begin{align}
&p_b(d_w, \theta_w) =\notag\\
& \frac{1}{1 \!+\! a \exp\left\{\!-\!b \left[\frac{180}{\pi}\sin^{\!-\!1}\left(\frac{d_w \sin(\theta_w)}{\sqrt{L^2 + d_w^2 - 2d_w L \cos(\theta_w)}}\right)\right]\!-\!a\right\}},
\end{align}
and
\begin{align}\label{bar_fw}
&\bar{f}(d_w) = \xi_{{\textrm{L}}} \left[d_w \!-\! L\cos(\theta_w)\right] +p_b(d_w, \theta_w)\sin(\theta_w)\notag\\
&~~~~\times\frac{180 a b}{\pi}\frac{L[L\!-\! d_w \cos(\theta_w)]}{|L\!-\! d_w \cos(\theta_w)|}\notag\\
&\times \exp\left\{\!-\!b \left[\frac{180}{\pi}\sin^{\!-\!1}\left(\frac{d_w \sin(\theta_w)}{\sqrt{L^2 \!+\! d_w^2 \!-\! 2d_w L \cos(\theta_w)}}\right)\right]\!-\!a\right\}\notag\\
&=\xi_{{\textrm{L}}} \left[d_w \!-\! L\cos(\theta_w)\right] +\sin(\theta_w)\notag\\
&~~~~\times\frac{180 b}{\pi}\frac{L[L\!-\! d_w \cos(\theta_w)]}{|L\!-\! d_w \cos(\theta_w)|}\left[1\!-\! p_b(d_w, \theta_w)\right].
\end{align}
Considering $p_b(d_w, \theta_w) >0$ and $L^2 + d_w^2 - 2d_w L \cos(\theta_w) = d_b^2>0$, following \eqref{f_dw_simple} we note that the solution of $d_w$ to ${\partial f(d_w, \theta_w)}/{\partial d_w} = 0$ is the same as the solution to $\bar{f}(d_w) = 0$. We also note that we have $\xi_{{\textrm{L}}}<0$, $\sin(\theta_w) >0$ due to $0 \leq \theta_w\leq \frac{\pi}{2}$, and $p_b(d_w, \theta_w) <1$ in \eqref{bar_fw}. As such, as per \eqref{bar_fw} we have $\bar{f}(d_w) < 0$ when $d_w > L/\cos(\theta_w)$, which leads to the conclusion of having ${\partial f(d_w, \theta_w)}/{\partial d_w} < 0$ for $d_w > L/\cos(\theta_w)$ according to \eqref{f_dw_simple}. Likewise, we have another conclusion of having ${\partial f(d_w, \theta_w)}/{\partial d_w} > 0$ for $d_w < L\cos(\theta_w)$. These two conclusions lead to that the solution to ${\partial f(d_w, \theta_w)}/{\partial d_w} = 0$ is in the closed interval $[L\cos(\theta_w), L/\cos(\theta_w)]$. In this closed interval, following \eqref{bar_fw} the function $\bar{f}(d_w) $ can be simplified as
\begin{align}\label{bar_fw_simple}
&\bar{f}(d_w) \!=\!\xi_{{\textrm{L}}} \left[d_w \!\!-\!\! L\cos(\theta_w)\right] +\frac{180 b L \sin(\theta_w)}{\pi}\left[1\!\!-\!\! p_b(d_w, \theta_w)\right].
\end{align}

We next prove that this solution is unique. To this end, we first derive the first derivative of $p_b(d_w, \theta_w)$ with respect to $d_w$ as
\begin{align}\label{p_b_dw_1st}
&\frac{\partial p_b(d_w, \theta_w) }{\partial d_w} \!\!=\!\! p_b^2(d_w, \theta_w)\left[L^2 \!+\! d_w^2 \!-\! 2d_w L \cos(\theta_w)\right]^{-1}\frac{180}{\pi}ab  \notag\\
&\times\exp\left\{\!-\!b \left[\frac{180}{\pi}\sin^{\!-\!1}\left(\frac{d_w \sin(\theta_w)}{\sqrt{L^2 + d_w^2 - 2d_w L \cos(\theta_w)}}\right)\right]\!\!-\!\!a\right\}\notag\\
& ~~~~\times\frac{ \sin(\theta_w)L[L- d_w \cos(\theta_w)]}{|L- d_w \cos(\theta_w)|}.
\end{align}
Following \eqref{p_b_dw_1st}, we note that ${\partial p_b(d_w, \theta_w) }/{\partial d_w}>0$ in the closed interval $[L\cos(\theta_w), L/\cos(\theta_w)]$. Noting $\xi_{{\textrm{L}}}<0$ and following \eqref{bar_fw_simple}, we can see that $\bar{f}(d_w)$ monotonically decreases with $d_w$ in the interval $[L\cos(\theta_w), L/\cos(\theta_w)]$, which leads to the fact that the solution to $\bar{f}(d_w) = 0$ is unique and thus the solution to ${\partial f(d_w, \theta_w)}/{\partial d_w} = 0$ is also unique in the closed interval $[L\cos(\theta_w), L/\cos(\theta_w)]$. This completes the proof of this lemma.
\end{IEEEproof}

\section*{Appendix B: Proof of Lemma~\ref{lemma2}}
The first derivative of $f(d_w, \theta_w)$ with respect to $\theta_w$ is derived as
\begin{align}\label{f_thetaw_1st}
&\frac{\partial f(d_w, \theta_w)}{\partial \theta_w}
=\frac{p_b(d_w, \theta_w)\xi_{{\textrm{L}}}d_w L\sin(\theta_w)}{\left[L^2 \!+\! d_w^2 \!-\! 2d_w L \cos(\theta_w)\right]^{1-\frac{\xi_{{\textrm{L}}}}{2}}}+p_b^2(d_w, \theta_w)\notag\\
&~~~~\times\left[L^2 \!+\! d_w^2 \!-\! 2d_w L \cos(\theta_w)\right]^{\frac{\xi_{{\textrm{L}}}}{2}}\frac{180}{\pi}ab \notag\\
&\!\times\!\exp\left\{\!-\!b \left[\frac{180}{\pi}\sin^{\!-\!1}\left(\frac{d_w \sin(\theta_w)}{\sqrt{L^2 + d_w^2 - 2d_w L \cos(\theta_w)}}\right)\right]\!\!-\!\!a\right\}\notag\\
&\!\times\! \frac{d_w}{|L\!-\! d_w \cos(\theta_w)|} \frac{(L^2 + d_w^2)\cos(\theta_w) \!-\! d_w L [1+\cos^2(\theta_w)]}{L^2 + d_w^2 - 2d_w L \cos(\theta_w)}\\
&\!=\!p_b(d_w, \theta_w)d_w\left[L^2 \!+\! d_w^2 \!\!-\!\! 2d_w L \cos(\theta_w)\right]^{\frac{\xi_{{\textrm{L}}}}{2}\!-\!1}\!\times\! \bar{g}(\theta_w) ,
\end{align}
where
\begin{align}\label{bar_g_thetaw}
&\bar{g}(\theta_w) = \xi_{{\textrm{L}}} L \sin(\theta_w) +p_b(d_w, \theta_w)\frac{180 a b}{\pi}\notag\\
&~~~~\times\frac{(L^2 + d_w^2)\cos(\theta_w) - d_w L [1+\cos^2(\theta_w)]}{|L\!-\! d_w \cos(\theta_w)|}\notag\\
&\!\times\! \exp\left\{\!-\!b \left[\frac{180}{\pi}\sin^{\!-\!1}\left(\frac{d_w \sin(\theta_w)}{\sqrt{L^2 + d_w^2 - 2d_w L \cos(\theta_w)}}\right)\right]\!\!-\!\!a\right\}\notag\\
&\!=\!\xi_{{\textrm{L}}} L \sin(\theta_w)  \!\!+\!\!\frac{180 b}{\pi}\frac{\mathcal{G}(\theta_w)}{|L\!-\! d_w \cos(\theta_w)|}\left[1\!-\! p_b(d_w, \theta_w)\right].
\end{align}
Considering $p_b(d_w, \theta_w) >0$, $L^2 + d_w^2 - 2d_w L \cos(\theta_w) = d_b^2>0$, and $d_w>0$, following \eqref{f_thetaw_1st} we note that the solution of $\theta_w$ to ${\partial f(d_w, \theta_w)}/{\partial \theta_w} = 0$ is the same as the solution of $d_w$ to $\bar{g}(\theta_w) = 0$. We next determine the value range of the optimal $\theta_w$. Following \eqref{G_thetaw_de}, the first derivative of $\mathcal{G}(\theta_w)$ with respect to $\theta_w$ is derived as
\begin{align}\label{G_thetaw_1t}
\frac{\partial \mathcal{G}(\theta_w)}{\partial \theta_w} &= -(L^2 + d_w^2)\sin(\theta_w) + 2 d_w L \sin(\theta_w) \cos(\theta_w)\notag\\
&= -\sin(\theta_w) d_b^2 <0,
\end{align}
which indicates that $\mathcal{G}(\theta_w)$ monotonically decreases with $\theta_w$. We note that $\mathcal{G}(\arccos (d_w/L)) =0$ when $d_w \leq L$ holds, and thus following \eqref{G_thetaw_de} we have $\mathcal{G}(\theta_w) <0$ for $\theta_w \geq \arccos (d_w/L)$ and $\mathcal{G}(\theta_w) >0$ for $\theta_w < \arccos (d_w/L)$. Following \eqref{bar_g_thetaw} and noting $\xi_{{\textrm{L}}}  <0$, $\sin(\theta_w) >0$, and $1 - p_b(d_w, \theta_w) >0$, we know that $\bar{g}(\theta_w) = 0$ requires $\mathcal{G}(\theta_w)  >0$, which requires $\theta_w < \arccos (d_w/L)$. As such, we can conclude that the solution to $\bar{g}(\theta_w) = 0$ is within the closed interval $[0, \arccos (d_w/L)]$.

In the following, we prove that the solution is unique within the interval $[0, \arccos (d_w/L)]$. To this end, we derive the first derivative of $p_b(d_w, \theta_w)$ with respect to $\theta_w$ as
\begin{align}\label{p_thetaw_1st}
&\frac{\partial p_b(d_w, \theta_w)}{\partial \theta_w} =p_b^2(d_w, \theta_w)\frac{180}{\pi}ab \exp\left\{\!-\!b \left[\frac{180}{\pi}\right.\right.\notag\\
&~~~~\times \left.\left.\sin^{\!-\!1}\left(\frac{d_w \sin(\theta_w)}{\sqrt{L^2 + d_w^2 - 2d_w L \cos(\theta_w)}}\right)\right]\!-\!a\right\}\notag\\
&~~~~\times \frac{d_w}{|L\!-\! d_w \cos(\theta_w)|} \frac{\mathcal{G}(\theta_w) }{L^2 + d_w^2 - 2d_w L \cos(\theta_w)}.
\end{align}
Considering the values of $\mathcal{G}(\theta_w)$ discussed below \eqref{G_thetaw_1t}, following \eqref{p_thetaw_1st} we have
${\partial p_b(d_w, \theta_w)}/{\partial \theta_w} >0$ for $\theta_w \leq \arccos (d_w/L)$, which demonstrates that $p_b(d_w, \theta_w)$ monotonically increases with $\theta_w$ for $\theta_w \leq \arccos (d_w/L)$ and thus $1- p_b(d_w, \theta_w)>0$ monotonically decreases with $\theta_w$ for $\theta_w \leq \arccos (d_w/L)$. We note that $\xi_{{\textrm{L}}}\sin(\theta_w)$ monotonically decreases with $\theta_w$ due to $\xi_{{\textrm{L}}}<0$. In addition, as per \eqref{G_thetaw_1t} and its following discussions, we know that $\mathcal{G}(\theta_w)>0$ and $\mathcal{G}(\theta_w)$ monotonically decreases with $\theta_w$ for $\theta_w < \arccos(d_w/L)$. Furthermore, $1/|L - d_w \cos(\theta_w)|$ monotonically decreases with $\theta_w$ for $L - d_w \cos(\theta_w) >0$. Therefore, following \eqref{bar_g_thetaw}, we can conclude that $\bar{g}(\theta_w)$ monotonically decreases with $\theta_w$ within the closed interval $[0, \arccos(d_w/L)]$ when $L - d_w \cos(\theta_w) >0$, which leads to that the solution to $\bar{g}(\theta_w) = 0$ is unique within the interval $[0, \arccos(d_w/L)]$ for $L - d_w \cos(\theta_w) >0$.

We recall that we have $\mathcal{G}(\theta_w) <0$ for $\theta_w \geq \arccos (d_w/L)$ according to \eqref{G_thetaw_1t} and its following discussions. Then, following \eqref{bar_g_thetaw}, and noting $\xi_{{\textrm{L}}}  <0$ and $1 - p_b(d_w, \theta_w) >0$, we have $\bar{g}(\theta_w) <0$ and thus ${\partial f(d_w, \theta_w)}/{\partial \theta_w} <0$ for $\theta_w \geq \arccos(d_w/L)$, which indicates that $f(d_w, \theta_w)$ monotonically decreases with $\theta_w$ for $\arccos(d_w/L) \leq \theta_w \leq \frac{\pi}{2}$ and thus completes the proof.

\bibliographystyle{IEEEtran}

\balance

\end{document}